\PassOptionsToPackage{unicode}{hyperref}
\PassOptionsToPackage{hyphens}{url}
\PassOptionsToPackage{dvipsnames,svgnames,x11names}{xcolor}
\documentclass[
  12pt]{article}

\usepackage{tikz}
\usepackage{hyperref}
\usepackage{tikz-network}
\usepackage{changepage}
\usepackage{tabularx}
\usepackage{amsmath,amssymb}
\usepackage{iftex}
\ifPDFTeX
  \usepackage[T1]{fontenc}
  \usepackage[utf8]{inputenc}
  \usepackage{textcomp} 
\else 
  \usepackage{unicode-math}
  \defaultfontfeatures{Scale=MatchLowercase}
  \defaultfontfeatures[\rmfamily]{Ligatures=TeX,Scale=1}
\fi
\usepackage{lmodern}
\ifPDFTeX\else  
\fi
\IfFileExists{upquote.sty}{\usepackage{upquote}}{}
\IfFileExists{microtype.sty}{
  \usepackage[]{microtype}
  \UseMicrotypeSet[protrusion]{basicmath} 
}{}
\makeatletter
\@ifundefined{KOMAClassName}{
  \IfFileExists{parskip.sty}{%
    \usepackage{parskip}
  }{
    \setlength{\parindent}{0pt}
    \setlength{\parskip}{6pt plus 2pt minus 1pt}}
}{
  \KOMAoptions{parskip=half}}
\makeatother
\usepackage{xcolor}
\makeatletter
\ifx\paragraph\undefined\else
  \let\oldparagraph\paragraph
  \renewcommand{\paragraph}{
    \@ifstar
      \xxxParagraphStar
      \xxxParagraphNoStar
  }
  \newcommand{\xxxParagraphStar}[1]{\oldparagraph*{#1}\mbox{}}
  \newcommand{\xxxParagraphNoStar}[1]{\oldparagraph{#1}\mbox{}}
\fi
\ifx\subparagraph\undefined\else
  \let\oldsubparagraph\subparagraph
  \renewcommand{\subparagraph}{
    \@ifstar
      \xxxSubParagraphStar
      \xxxSubParagraphNoStar
  }
  \newcommand{\xxxSubParagraphStar}[1]{\oldsubparagraph*{#1}\mbox{}}
  \newcommand{\xxxSubParagraphNoStar}[1]{\oldsubparagraph{#1}\mbox{}}
\fi

\makeatother
\newcommand{\bd}{\boldsymbol} 
\newcommand{\mb}{\mathbf} 

\newcommand{\yi}{y_{hij}}
\newcommand{\ti}{t_{hij}}
\newcommand{\di}{d_{hij}}

\usepackage{ulem}

\newcommand{\be}{\begin{equation}}
\newcommand{\ee}{\end{equation}}

\usepackage{longtable,booktabs,array}
\usepackage{calc} 
\usepackage{etoolbox}
\makeatletter
\patchcmd\longtable{\par}{\if@noskipsec\mbox{}\fi\par}{}{}
\makeatother
\IfFileExists{footnotehyper.sty}{\usepackage{footnotehyper}}{\usepackage{footnote}}
\makesavenoteenv{longtable}
\usepackage{graphicx}
\makeatletter
\def\maxwidth{\ifdim\Gin@nat@width>\linewidth\linewidth\else\Gin@nat@width\fi}
\def\maxheight{\ifdim\Gin@nat@height>\textheight\textheight\else\Gin@nat@height\fi}
\makeatother
\setkeys{Gin}{width=\maxwidth,height=\maxheight,keepaspectratio}
\makeatletter
\def\fps@figure{htbp}
\makeatother

\makeatletter
\@ifpackageloaded{caption}{}{\usepackage{caption}}
\AtBeginDocument{%
\ifdefined\contentsname
  \renewcommand*\contentsname{Table of contents}
\else
  \newcommand\contentsname{Table of contents}
\fi
\ifdefined\listfigurename
  \renewcommand*\listfigurename{List of Figures}
\else
  \newcommand\listfigurename{List of Figures}
\fi
\ifdefined\listtablename
  \renewcommand*\listtablename{List of Tables}
\else
  \newcommand\listtablename{List of Tables}
\fi
\ifdefined\figurename
  \renewcommand*\figurename{Figure}
\else
  \newcommand\figurename{Figure}
\fi
\ifdefined\tablename
  \renewcommand*\tablename{Table}
\else
  \newcommand\tablename{Table}
\fi
}
\@ifpackageloaded{float}{}{\usepackage{float}}
\floatstyle{ruled}
\@ifundefined{c@chapter}{\newfloat{codelisting}{h}{lop}}{\newfloat{codelisting}{h}{lop}[chapter]}
\floatname{codelisting}{Listing}

\makeatother
\makeatletter
\@ifpackageloaded{caption}{}{\usepackage{caption}}
\@ifpackageloaded{subcaption}{}{\usepackage{subcaption}}
\makeatother

\usepackage{varwidth}
\newcommand{\ac}[1]{%
  \fbox{%
    \textcolor{magenta}{%
      \begin{varwidth}{\linewidth}%
        \textbf{Aaron's comment:} #1%
      \end{varwidth}%
    }%
  }%
}

\ifLuaTeX
  \usepackage{selnolig}  
\fi
\usepackage{natbib}
\usepackage{bookmark}

\IfFileExists{xurl.sty}{\usepackage{xurl}}{} 
\hypersetup{
  pdftitle={Title},
  pdfauthor={Author 1; Author 2},
  pdfkeywords={3 to 6 keywords, that do not appear in the title},
  colorlinks=true,
  linkcolor={blue},
  filecolor={Maroon},
  citecolor={Blue},
  urlcolor={magenta},
  pdfcreator={LaTeX via pandoc}}

\newcommand{\anon}{1}

\begin{document}

\def\spacingset#1{\renewcommand{\baselinestretch}%
{#1}\small\normalsize} \spacingset{1}


\if1\anon

\title{\bf A Bayesian Threshold-Aligned Joint Disease Progression Model (B-TAJ DPM) for Alzheimer's Disease}

\author{
Rong Wu\\
Department of Epidemiology and Biostatistics, \\
University of California, San Francisco
\and
Duygu Tosun\\
Department of Radiology and Biomedical Imaging, \\
University of California, San Francisco
\and
Isabella Hausle\\
Department of Radiology and Biomedical Imaging, \\
University of California, San Francisco
\and
Margo Heston\\
Department of Neurology, \\
University of California, San Francisco
\and
Aaron Wolfe Scheffler\\
Department of Epidemiology and Biostatistics, \\
University of California, San Francisco\\
for the Alzheimer’s Disease Neuroimaging Initiative*
  \thanks{
    Data used in preparation of this article were obtained from the Alzheimer’s Disease
Neuroimaging Initiative (ADNI) database (adni.loni.usc.edu). As such, the investigators
within the ADNI contributed to the design and implementation of ADNI and/or provided data
but did not participate in analysis or writing of this report. A complete listing of ADNI
investigators can be found at:
\url{http://adni.loni.usc.edu/wp-content/uploads/how_to_apply/ADNI_Acknowledgement_List.pdf}}
}

\maketitle

\fi

\if0\anon
{
  \bigskip
  \bigskip
  \bigskip
  \begin{center}
    {\LARGE\bf A Bayesian Threshold-Aligned Joint Disease Progression Model (B-TAJ DPM) for Alzheimer's Disease}
\end{center}
  \medskip
} \fi

\bigskip
\begin{abstract}
Alzheimer's disease is characterized by the progressive accumulation of amyloid-$\beta$ and tau followed years later by cognitive impairment. Despite this established motif, substantial subject-level variability exists in the age of pathological progression and the onset of cognitive symptoms. To understand the source of this variation, subjects must be aligned across heterogeneous disease timelines via frameworks that jointly model disease progression and time to cognitive impairment with reference to landmark positivity thresholds. Existing neurodegenerative disease progression models rely on restrictive parametric forms, fail to anchor disease timelines to positivity thresholds, and decouple biomarker trajectories from cognitive survival endpoints. To address these limitations, we introduce the Bayesian Threshold-Aligned Joint Disease Progression Model (B-TAJ DPM). This generative, semi-parametric framework models multivariate disease progression trajectories over latent disease timelines anchored at landmark positivity thresholds. Crucially, the framework integrates a survival model to link pathological progression to cognitive impairment. Posterior inference and posterior predictions for unseen subjects are carried out in open-source software. Simulation studies demonstrate excellent estimation accuracy and interval coverage. When applied to Alzheimer's Disease Neuroimaging Initiative data, B-TAJ DPM characterizes non-linear progression patterns, quantifies subject-level variation in positivity age, and reveals links between age of tau positivity and acceleration of cognitive impairment.
\end{abstract}

\noindent%
\vfill

\newpage
\spacingset{1.8} 

\section{Introduction}\label{intro}

Alzheimer's disease (AD), a leading cause of global disability and mortality, is marked by the accumulation of amyloid-$\beta$ plaques and tau neurofibrillary tangles in the brain, signaling pathological processes that begin years before the onset of cognitive impairment \citep{Jack2024}. It is generally agreed that amyloid deposition occurs throughout the brain several years before tau begins to accumulate in the entorhinal, parahippocampus, and amygdala regions \citep{duygu2025}. Despite this accepted motif, there is tremendous variability among subjects in age of amyloid and tau presentation, the time between pathological progression, and the lag between pathological disease and onset of cognitive impairment. A deeper understanding of this subject-level variability is critical to understand the etiology of AD and plan future interventional trials.

\begin{figure}[H]
\centering\includegraphics[width=0.9\textwidth]{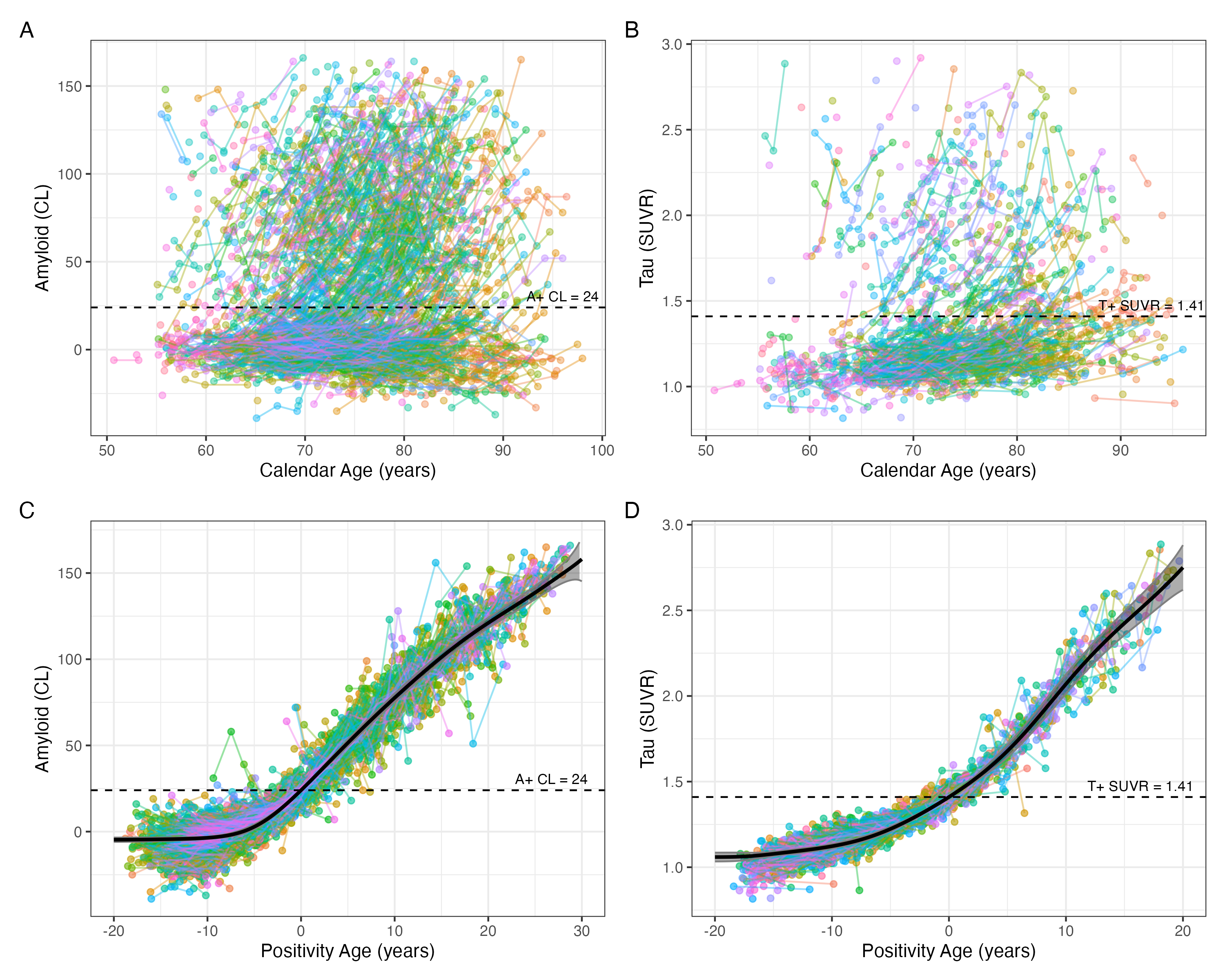}
    \caption{Observed subject-level trajectories for (A) global cortical amyloid (CL) and (B) mesial-temporal tau (SUVR) aligned by calendar age (years). Positivity thresholds for amyloid (A+ = 24 CL) and tau (T+ = 1.41 SUVR) are displayed as horizontal dashed lines. The same subject-level trajectories are shown for amyloid (C) and tau (D) aligned by estimated positivity age (years) from the proposed B-TAJ DPM procedure. Solid black lines are posterior means of the progression trajectories for each biomarker accompanied by gray bands representing 95$\%$ credible intervals.}
\label{fig:study_dta}
\end{figure}

To achieve this, clinical studies of AD use positron emission tomography (PET) imaging to measure amyloid and tau via brain scans collected on subjects longitudinally over short durations of time. For each subject, the resulting data is a quantitative measure of amyloid and tau burden indexed by calendar age (Figure \ref{fig:study_dta} (A,B)). These quantitative PET measurements can be compared against clinical thresholds to determine whether a patient has attained biomarker levels above those in healthy subjects, i.e. whether a patient has crossed a positivity threshold \citep{Heston2025, duygu2025}. In fact, positivity thresholds act as reference points for patients across disease stages and can be used to quantify key subject-level quantities such as age of amyloid (A+) and tau positivity (T+), time between A+ and T+, and time from A+ or T+ to cognitive impairment. In this way, subject-specific disease timelines can be constructed to better understand variation in disease progression. 

Several challenges emerge when estimating subject-specific positivity timelines. First, although observations on any individual may span several years, they do not cover the full continuum of disease progression and it is rare to observe the transition to positivity. Therefore, a disease progression template is needed against which subjects can be aligned in order to predict their individual age of positivity. Second, because subjects experience disease onset at different calendar ages, estimation of a disease progression template requires modeling a latent disease timeline that is never observed, e.g. positivity age. To address these challenges, a rich literature in neurodegenerative disease progression models (ND-DPMs) has emerged which will be discussed thoroughly in Section \ref{subsec- 1 related_work}. However, existing ND-DPM approaches suffer from one or more of the following weaknesses: disease progression templates are constrained to reductive parametric forms, positivity thresholds are not considered as anchor points when estimating latent disease timeline, estimation procedures are heuristic and lack principled inference, and pathological progression is detached from time until cognitive impairment rather than considering joint longitudinal and survival models to link the two processes. As such, there is no joint framework that simultaneously models multiple PET biomarkers along a positivity timeline using semi-parametric methods while linking that timeline to benchmarks of cognitive impairment. As a result, current models cannot estimate expressive progression templates along an interpretable positivity age timelines while pooling information across pathological and cognitive domains, all while providing principled uncertainty quantification. Collectively, this limits our ability to properly model and account for subject-level variation in disease progression.  

\subsection{Scientific objectives and our proposed Bayesian framework} \label{subsec-1 obj}
 
We introduce a Bayesian framework to achieve the following scientific objectives. First, for each PET biomarker, we will estimate a disease progression trajectory without imposing parametric assumptions. The trajectory must be monotonic since AD is progressive with the constraint that the latent disease timeline is anchored by a prespecified positivity threshold. Second, we will estimate the age of A+ and T+ for each subject while accounting for subject-level risk factors and within-subject correlations in order to partition variation in predicted age of positivity. Third, we will model associations between subject-level risk factors and time from T+ to cognitive impairment jointly with disease progression trajectories, allowing information from pathological and cognitive processes to inform each other. Fourth, we will provide principled uncertainty quantification for each model parameter to improve scientific interpretation of model findings and post-estimation inference.  

To achieve these objectives, we introduce a Bayesian Threshold-Aligned Joint Disease Progression Model (B-TAJ DPM) for AD. We adopt a Bayesian framework to facilitate principled uncertainty quantification in addition to its innate ability to recover latent parameters via hierarchical model specifications. For each PET biomarker, we assume a population-level disease progression trajectory whose log transformed derivative is projected onto a B-spline basis with a prior structure that allows for adaptive smoothing. The derivative is integrated over a latent disease timeline with the constraint that the disease timeline has a value of zero when a prespecified positivity threshold is crossed. Correlated subject-specific random linear shifts align each subject to the population-level disease progression trajectory. We define an accelerated failure time model for the time from T+ to cognitive impairment, linking pathological and cognitive information via a common generative framework. For both age of A+, age of T+, and time from T+ to cognitive impairment, hidden linear models are embedded to model the association with risk factors of interest. The result is a principled Bayesian framework that simultaneously models and aligns short term multivariate longitudinal PET measurements along a positivity timeline while jointly modeling time to cognitive impairment. The model is described in detail in Section \ref{sec-meth}.
    
\subsection{Novelty of the proposed work}\label{subsec- 1 intro_novelty}

Our proposed B-TAJ DPM framework provides several innovations. To our knowledge, this is the first generative, semi-parametric model to estimate monotonic multivariate disease progression trajectories where the latent disease timeline is anchored by a prespecified, outcome-specific positivity threshold. Next, by jointly modeling the multivariate outcome trajectories with time from positivity to cognitive impairment, we propose the first model to perform curve alignment and survival analysis in the context of monotonic progression models. Finally, we develop an estimation framework that not only provides computationally efficient estimation of posterior distributions of all model parameters in this novel setting but also allows for posterior predictions for unseen subjects via easily implementable statistical software. Importantly, this facilitates the dissemination of cutting edge models to clinical researchers interrogating subject-level variation in AD progression.

\subsection{Related work in neurodegenerative disease progression models (ND-DPMs)}\label{subsec- 1 related_work}

Prior work on ND-DPMs draws on a broader statistical literature that addresses (1) monotonicity, (2) curve alignment, and (3) joint longitudinal and survival models. We begin by reviewing the statistical literature related to these topics and then focus closely on a selection of ND-DPMs most related to our scientific objectives and inferential goals. With regard to monotonicity, a broad class of statistical methods has been proposed to enforce shape constraints via customized basis functions, projections, and constrained optimization \citep{ramsay1988, jones2009, maad2020, meyer2008, abraham2015,lin2014}. Curve alignment has been considered from two perspectives. First, functional data analysis (FDA) methods decompose subject-level variation into amplitude and phase variation, where the latter can be removed via nonlinear transformations of the functional domain in order to provide estimates of common shape functions \citep{telesca2008, wrobel2019, Matuk2022, Raket2014, Marron2014}. Second, self-modeling regression frameworks often assume a common semi-parametric shape function and shift subjects to a common domain via linear shifts \citep{kneip1988convergence,lawton1972self,ladd2000self, brumback2004self, beath2007infant, lindstrom1995self, wang2003shape, altman2004self, coull2004self}. FDA methods often focus on densely observed data whereas self modeling regression is better suited for sparsely observed data. Lastly, joint longitudinal and survival analysis has received considerable attention \citep{Wang2025} but only one work has specifically considered joint modeling within the context of recovering a latent disease timeline via alignment \citep{Wang2023}. This approach is innovative but the disease progression trajectory lacks monotonic constraints and does not consider prespecified positivity thresholds. 

We now focus specifically on ND-DPMs which address monotonicity and curve alignment simultaneously. In a sense, the simplest ND-DPMs assume the disease progression template has a parametric form, either linear or nonlinear, with subjects aligned via random linear shifts for univariate \citep{Raket2020_alz, Jedynak2012, Oxtoby2014} and multivariate trajectories \citep{Kuhnel2021, Poulet2023, Staffaroni2022, Donohue2014, Therneau2021, Koval2021}. These models were foundational in characterizing ND progression but enforce restrictive parametric assumptions on the assumed disease trajectory risking misspecification. To address this limitation, semi-parametric and nonparametric techniques including constrained Gaussian processes \citep{Abinader2020, Lorenzi2019}, basis expansions \citep{bilgel2016}, and piecewise linear methods \citep{Young2018, Young2021} have emerged to estimate monotonic disease progression trajectories. While these methods provide flexible template functions for alignment, they lack a clear framework to anchor disease progression against prespecified positivity thresholds. Existing methods that come closest to meeting our scientific objectives are the Sampled Iterative Local Approximation (SILA) method \citep{Betthauser2022} and Temporal Integration of Rate Accumulation method (TIRA) method \citep{Budgeon2017, Schindler2021}. Both methods estimate the local rate of change of subject-level trajectories, map their relationship to the magnitude of the biomarker in question, and then solve the resulting ODE to obtain a disease progression trajectory indexed by time since a fixed positivity threshold. Both approaches are widely used to model subject-level variation in age of A+ and T+ but rely on heuristic estimation procedures that do not allow multivariate outcome trajectories, cannot directly incorporate covariate effects, do not jointly model time-to-event processes, and can only begin to provide uncertainty quantification via ad hoc bootstrap procedures. Thus, despite a robust literature on ND-DPMs, no single model allows us to adequately address our scientific objectives. 


The remainder of the paper is structured as follows. Section 2 describes our motivating data from the Alzheimer's Disease Neuroimaging Initiative study \citep{Petersen2010}, Section 3 formally introduces our proposed B-TAJ DPM, Section 4 evaluates our proposed methodology via an extensive simulation study, Section 5 applies B-TAJ DPM to the ADNI data, and Section 6 concludes with a discussion and directions for future work. 

\section{ADNI Study}\label{sec:data_description}

Data used in the preparation of this article were obtained from the Alzheimer’s Disease Neuroimaging Initiative (ADNI) database (adni.loni.usc.edu). The primary objective of ADNI has been to determine whether serial magnetic resonance imaging (MRI), positron emission tomography (PET), other biological markers, and clinical and neuropsychological assessments can be integrated to track the progression of mild cognitive impairment (MCI) and early AD. To achieve our scientific objectives described in Section \ref{subsec-1 obj}, we will consider two pathological markers (amyloid and tau), 
one cognitive outcome and several subject-level risk factors (sex, college education and APOE-$\varepsilon$4 genotype). 

\begin{table}[htbp]
\centering
\caption{ADNI study sample characteristics stratified by subjects with amyloid, tau, and the subset of subjects with T+ prior to mild cognitive impairment.}
\label{tab:adni_table1}

\footnotesize
\setlength{\tabcolsep}{3pt}
\renewcommand{\arraystretch}{1.05}
\scriptsize
\setlength{\tabcolsep}{2.5pt}
\renewcommand{\arraystretch}{0.6}
\begin{tabularx}{\linewidth}{
>{\raggedright\arraybackslash}p{2.8cm}
>{\raggedright\arraybackslash}p{2.8cm}
>{\centering\arraybackslash}X
>{\centering\arraybackslash}X
>{\centering\arraybackslash}X
}
\toprule

& & \textbf{Amyloid} & \textbf{Tau} & \textbf{T+ to CDR $\geq$ 0.5} \\

\midrule

\textbf{n} & & 1900 & 991 & 293 \\
& & (962 A+) & (309 T+) & \\

\midrule

\textbf{Sex (\%)} 
& Male   & 944 (49.7) & 464 (46.8) & 128 (43.7) \\
& Female & 956 (50.3) & 527 (53.2) & 165 (56.3) \\

\midrule

\textbf{Race (\%)} 
& Asian & 60 (3.2) & 32 (3.3) & 6 (2.1) \\
& Black or Afr Am & 170 (9.0) & 117 (11.9) & 21 (7.2) \\
& $> 1$ race & 25 (1.3) & 15 (1.5) & 7 (2.4) \\
& Unknown & 16 (0.8) & 9 (0.9) & 3 (1.0) \\
& White & 1613 (85.3) & 808 (82.1) & 254 (87.3) \\
& Other & 7 (0.4) & 3 (0.3) & -- \\

\midrule

\textbf{APOE (\%)} 
& 2/2 & 6 (0.3) & 2 (0.2) & 1 (0.3) \\
& 2/3 & 154 (8.1) & 93 (9.4) & 9 (3.1) \\
& 2/4 & 41 (2.2) & 21 (2.1) & 7 (2.4) \\
& 3/3 & 929 (48.9) & 507 (51.2) & 95 (32.4) \\
& 3/4 & 614 (32.3) & 301 (30.4) & 136 (46.4) \\
& 4/4 & 156 (8.2) & 67 (6.8) & 45 (15.4) \\

\midrule

\textbf{Baseline Dx (\%)} 
& Unimpaired & 763 (43.8) & 466 (56.8) & 72 (28.8) \\
& MCI & 729 (41.9) & 259 (31.5) & 106 (42.4) \\
& Dementia & 249 (14.3) & 96 (11.7) & 72 (28.8) \\

\midrule

\textbf{Final Dx (\%)} 
& Unimpaired & 638 (40.7) & 444 (56.4) & 52 (22.7) \\
& MCI & 573 (36.6) & 232 (29.5) & 90 (39.3) \\
& Dementia & 355 (22.7) & 111 (14.1) & 87 (38.0) \\

\midrule

\textbf{Age at First Scan}
& Mean (SD) & 72.52 (7.69) & 73.07 (8.17) & 75.78 (7.76) \\

\textbf{Follow-up Years}
& Mean (SD) & 2.85 (3.35) & 2.04 (2.53) & 1.85 (2.34) \\

\midrule

\textbf{Scans (\%)} 
& 1 & 794 (41.8) & 488 (49.2) & 139 (47.4) \\
& 2 & 487 (25.6) & 254 (25.6) & 67 (22.9) \\
& 3 & 300 (15.8) & 160 (16.1) & 53 (18.1) \\
& 4 & 154 (8.1) & 61 (6.2) & 21 (7.2) \\
& 5 & 85 (4.5) & 25 (2.5) & 12 (4.1) \\
& 6 & 64 (3.4) & 1 (0.1) & 1 (0.3) \\
& 7 & 16 (0.8) & 2 (0.2) & 0 (0.0) \\

\bottomrule
\end{tabularx}
\end{table}
As displayed in Figure~\ref{fig:study_dta}(A,B), the pathological markers consist of longitudinal PET measurements of amyloid and tau pathology. Global cortical amyloid burden is quantified using Centiloid (CL) values derived from $^{18}\text{F}$-florbetapir (FBP) or $^{18}\text{F}$-florbetaben (FBB) scans, normalized to a composite reference region as described in ADNI protocols \citep{Jagust2015}. Tau pathology is measured by the standardized uptake value ratio (SUVR) from $^{18}\text{F}$-flortaucipir (FTP) imaging within a mesial-temporal meta-region of interest (ROI), normalized to the inferior cerebellar grey matter. Thresholds for PET biomarkers (A+ = 24 CL, T+ = 1.41 SUVR) are determined as described in \citet{duygu2025}. As tabulated in Table~\ref{tab:adni_table1}, $n = 1900$ subjects had amyloid measurements with 962 falling above the amyloid positivity threshold. A subset of $n = 991$ subjects had tau measurements with 309 falling above the tau positivity threshold. Each subject contributed 1 to 7 scans with the average age of first scan for amyloid and tau of 72.5 and 73.1 years old and an average follow-up time of 2.9 and 2.0 years, respectively.

Cognitive outcomes are measured by Clinical Dementia Rating (CDR) assessed at baseline and follow-up visits with mild cognitive impairment defined as a CDR $\geq$ 0.5.  For the purpose of modeling time from T+ to cognitive impairment, we restrict our analysis to the $n = 293$ subjects who became T+ prior or concurrent to the first observation of cognitive impairment. This results in 16 subjects being excluded from the survival analysis. This is required since we assume T+ precedes cognitive impairment. Subjects where this ordering is clearly violated may have their impairment driven by other co-pathologies and thus are not included in the survival analysis. 
Other subject-specific risk factors include 
biological sex, educational status, and APOE-$\varepsilon$4 carriage status which have received prior attention in the study of amyloid and tau progression and time from T+ to cognitive impairment \citep{Betthauser2022, Heston2025,duygu2025}. 

\section{Methods}\label{sec-meth}

\subsection{Bayesian hierarchical model}\label{subsec-bhm}

We now formally introduce our proposed Bayesian Threshold-Aligned Joint Disease Progression Model (B-TAJ DPM), a hierarchical Bayesian framework that simultaneously models and aligns short term multivariate longitudinal pathological markers while jointly modeling time to cognitive impairment.

\subsubsection{Pathological marker model}\label{sub2sec-biom}

We consider a set of $H$ pathological biomarkers denoted by $\mathcal{D} = \{\mathcal{Y}_1, \dots, \mathcal{Y}_H\}$. For each biomarker $h \in \{1, \dots, H\}$, let $\mathcal{Y}_h = \{(t_{hij}, y_{hij}) : i \in \mathcal{I}_h,\; j = 1, \dots, J_{hi}\}$ denote the set of observed longitudinal measurements. Here, $\mathcal{I}_h$ represents the index set of subjects with observations for the $h$-th biomarker, $t_{hij}$ is the $j$-th observed calendar age for subject $i$ on marker $h$, $y_{hij}$ is the corresponding biomarker measurement value, and $J_{hi}$ is the total number of observations for subject $i$ on marker $h$. Let $\mathcal{I} = \bigcup_{h=1}^H \mathcal{I}_h$ denote the unique set of all subjects in the study, where $|\mathcal{I}|$ represents the total sample size. Finally, let $\bd{x}_i \in \mathbb{R}^p$ denote a vector of static subject-level covariates for individual $i$. For the remainder of this section, we assume $H = 2$ as in our real data although our approach can easily be extended to higher dimensions.

Although the biomarker measurements are indexed by calendar age $t_{hij}$, 
disease progression is not synchronized across individuals and biomarkers. 
To account for this heterogeneity, we assume that for the $h$-th biomarker, 
the $i$-th subject has a subject-specific biomarker time shift parameter $\alpha_{hi}$, 
representing the age at which the corresponding biomarker crosses a prespecified positivity threshold $\tau_h$ indicative of the onset of the underlying pathological process.
This allows for the definition of a biomarker specific latent disease timeline,
$d_{hij} = t_{hij} - \alpha_{hi}$,
which aligns subjects’ disease progression trajectories according to their subject-specific age of positivity for each biomarker. Furthermore, we define a set of specified domains for 
modeling the progression trajectories across markers 
by restricting different disease age scales 
$\di \in [L_h, U_h]$ 
to lie within a bounded interval, 
where $L_h$ and $U_h$ denote the lower and upper bounds of the latent disease age for marker $h$, respectively. The upper and lower bounds should be determined by domain knowledge. During estimation, we constrain 
$d_{hij}$ like in the interval
$\mathbb{I}(L_h < \di < U_h)$.

Our resulting Bayesian hierarchical biomarker model is defined as
\begin{gather}
    \yi = f_h(\di) + \epsilon_{hij},\  \epsilon_{hij} \sim  \mathcal{N}(0, \sigma^2_{\epsilon_h}), \\
    \di = \ti - \alpha_{hi},\  \bd\alpha_i \sim \mathcal{BVN}(\bd\mu_i, \mb\Sigma_\alpha ),
    \label{eq:alpha_h} 
    \\ \mu_{hi} = \bd{x}_i'\bd{\gamma}_h,\ \bd\gamma_{h} \sim \mathcal{N}\{
    (\mu_{h\gamma},\bd{0}_{p-1})', \text{diag}(a, b\bd{1}_{p-1})\},
\end{gather}
where $f_h(\di)$ denotes the disease progression trajectory for the $h$-th biomarker as a function of latent disease time $\di$, 
$\mathcal{BVN}(\cdot)$ is a bivariate normal distribution with mean $\bd\mu_i = (\mu_{1i}, \mu_{2i})'$,
$\mb\Sigma_\alpha  = 
\begin{bmatrix}
\sigma_{\alpha_1}^2 \quad \rho\sigma_{\alpha_1}\sigma_{\alpha_2} \\
\rho\sigma_{\alpha_1}\sigma_{\alpha_2} \quad \sigma_{\alpha_2}^2
\end{bmatrix}
$ is the $2 \times 2$ covariance matrix, where $\rho$ is the correlation coefficient 
and $\sigma_{\alpha_1}$ and $\sigma_{\alpha_2}$ are standard deviations for age of positivity, $\bd\gamma_h \in \mathbb{R}^p$ are regression coefficients with the first component corresponding to the intercept with mean $\mu_{h\gamma}$ and the remaining $p-1$ components
corresponding to covariate effects, 
$\text{diag}(a, b\bd{1}_{p-1})$ is the $p \times p$ covariance matrix, 
and $a$ and $b$ are scalars controlling the prior variance of $\bd\gamma_h$, and $\epsilon_{hij} \sim \mathcal{N}(0, \sigma^2_{\epsilon_h})$ are i.i.d.\ Gaussian noises. Finally, we place priors on the variance parameters as:
\[
\sigma_{\epsilon_h} \sim \mathcal{N}^+(0, 1),
\
\sigma_{\alpha_h} \sim \mathcal{N}^+(\zeta_h, \eta_h),
\ \rho \sim U(-1,1),
\]
where $\zeta_h$ and $\eta_h$ reflect prior knowledge regarding variation in the age of positivity for each biomarker.

\subsubsection{Disease progression trajectory}\label{sub2sec-dpm}
In this section, we introduce our model for the strictly monotonic increasing disease progression trajectory $f_h(d_h)$. To do so, we target the positive derivative of the disease progression trajectory over the latent positivity timeline on the log scale. This allows us to project the log derivative onto an unconstrained B-spline basis \citep{de1978practical} that flexibly captures disease progression patterns and produces a strictly monotonically increasing trajectory following exponentiation and numerical integration.
Specifically,
we represent the log derivative for each biomarkers progression trajectory using a spline basis defined on a set of knots 
$(\kappa_1, \kappa_2, \dots, \kappa_{K_0})$
partitioning the sampling interval into
$K_0+1$ sub-intervals.
We use a cubic B-spline basis defined on a bounded disease-time interval 
$[L_h, U_h]$ defined in Section \ref{sub2sec-biom}.
Let $\mb{B}_h(d_h) = \{\bd b_{h1}(d_h),\ldots, \bd b_{hK}(d_h)\}'$ denote the resulting cubic B-spline basis evaluated at latent disease time $d_h$,
where $\bd b_{hK}(d_h) \in \mathbb{R}^{\ell}$ is a B-spline basis function defined on a $\ell$-dimensional even grid,
and $K = K_0+2$ denotes the total number of basis functions.
The log derivative of the disease progression trajectory is represented as
$\mb{B}(d_h)'\bd{\theta}_h,$
where $\bd{\theta}_h \in \mathbb{R}^K$ are a vector of B-spline coefficients that characterize the trajectory for each biomarker
We model $f_h(\cdot)$ as follows:
\begin{gather}
    f_h(d_h) = \int_{L_h}^{U_h} \text{exp}\{ \mb{B}(d_h)'\bd{\theta}_h\}dd_h + C_h, \label{eq:int} \\
    \quad \theta_{hk} \sim \mathcal{N}(\theta_{h(k-1)}, \lambda_{hk}/\sigma^2_{\theta_{h}}),\quad k = 2,\ldots, K,\\ \nonumber
    \sigma_{\theta_h}^{2} \sim {\Gamma}(1e^{-3}, 1e^{-3}), \quad \lambda_{hk} \sim \text{Exp}(1),\quad k = 2,\ldots, K,
\end{gather}

with $\theta_{h1} \sim \mathcal{N}(-9, 0.1^2)$ corresponding to a flat rate of change early in the disease course,
$\sigma_{\theta_h}^{2}$ is a global precision parameter,
and $\lambda_{hk}$ is a local adaptive smoothing parameter that allows the shrinkage between adjacent spline coefficients to vary across $k$. 
Smaller values of \(\lambda_{hk}\) induce stronger local shrinkage of \(\theta_{hk}\) toward \(\theta_{h(k-1)}\), 
while larger values indicate greater local deviations in the progression trajectory. We find that an adaptive smoothing prior is necessary given there is more posterior uncertainty in the disease progression trajectory as it moves away from the positivity threshold.

The use of the exponential transformation ensures that the derivative $f_h'(\cdot)$ is strictly positive for all $t$, which in turn guarantees that the trajectory $f_h(\cdot)$ is monotone increasing. 
To align individuals by disease time while ensuring that all evaluations are performed on a bounded grid, 
we construct a discrete centered time grid $\{d_\ell\}_{\ell=1}^{n_{d}}$ with spacing $\Delta d$ and total points $n_d = (U_h - L_h)/\Delta d + 1$, 
where 
$d_\ell = (\ell-n_d/2)\Delta d$.
For each biomarker, the population-level trajectory is evaluated on the same fixed grid and $d_{hij}$ is mapped to the nearest grid index by
$\ell_{hij} = \text{round}[\min\{\max(1,\dfrac{d_{hij}}{\Delta d} + \dfrac{n_d}{2}),n_d\}]$.
Numerical integration is implemented over this finite disease-age grid 
$\{d_\ell\}_{\ell=1}^{n_{d}}$.
By modeling the log derivative rather than the trajectory directly, monotonicity is enforced implicitly and offers flexible future extensions for modeling the derivative in the unconstrained log space.  
Lastly, the constant $C_h$ is fixed to ensure that $f_h(0) = \tau_h$ which allows for natural definition of a positivity age of 0 at the threshold $\tau_h$. Specifically, $C_h$ is calculated deterministically following numeric integration of the derivative. 

\subsubsection{Accelerated Failure Time model}\label{sub2sec-aft}
Let $T_i$ be the failure time defined as time from T+ to cognitive impairment for the $i$-th subject,
and $C_i$ denote the time point of right-censoring. 
Only $r_i = \min\{T_i,C_i\}$ and 
the event indicator
$\delta_i = \mathbb{I}(T_i \leq C_i)$ are observed.
To quantify the relationship between $T_i$ and risk factors of interest, 
we propose an accelerated failure time (AFT) model \citep{wei1992accelerated} that considers both subject-specific covariates as well as estimated age of T+. 
A typical AFT model is written as:
$\log(T_i) = \bd z_i'\bd{\phi}_h +  \epsilon_{i}$,
where $\bd z_i'$ denotes a $p' \times 1$ vector of risk factors for the $i$-th subject,
$\bd{\phi}_h$ are corresponding coefficients, 
and $\epsilon_{i}$ is random noise.
While a range of AFT models could be employed, we adopt the Weibull distribution for $T$ in this application. To link the shift parameter $\alpha_{hi}$ for a given biomarker to the time-to-event outcome $T_i$,
we utilize a three-parameter Weibull distribution as the survival process. Let $T$ denote the calendar age at which the clinical event such as passing a certain threshold for a cognitive scale (e.g. CDR $\ge 0.5$) for a subject, and let
$\alpha$ denote the corresponding subject-specific latent onset time, the distribution for $T$ can be written as:
$T \mid \alpha, \lambda, \nu \sim \text{Weibull}_3(\nu, \lambda, \alpha)$, 
with density function:
\begin{gather}
f_{\text{Weibull}_3}(T \mid \nu,\lambda,\alpha) =
\dfrac{\nu}{\lambda}
\left(\dfrac{T - \alpha}{\lambda}\right)^{\nu - 1}
\exp \Big[-\left(\frac{T-\alpha}{\lambda}\right)^{\nu} \Big], \label{eq:weib}
\ T>\alpha,
\end{gather}
where $\nu > 0$ is the shape parameter, $\lambda > 0$ is the scale parameter, 
and $\alpha$ acts as a shift parameter.
We adopt an AFT model under Equation \ref{eq:weib} as follows:
\begin{gather}
    T_i \mid \nu_h,\lambda_{hi},\alpha_{hi} \sim \text{Weibull}_3(\nu_h, \lambda_{hi}, \alpha_{hi}),\quad \nu_h \sim \text{Flat}^+(0, \infty),\\
\lambda_{hi} = \exp(\bd{z}_i'\bd{\phi}_h), \quad \bd{\phi}_h \sim \mathcal{N}(\bd{0}, c\mb I_{p'}),
\end{gather}
where 
$\nu_h > 0$ is the shape parameter, 
$\lambda_{hi} > 0$ is the subject-specific scale parameter, 
$\alpha_{hi}$ represents the time of biomarker positivity,
$\bd z_i \in \mathbb{R}^{p'}$ denotes a vector of covariates specific to the survival process 
with corresponding coefficients $\bd \phi_h$,
$\mb{I}_{p'}$ is the $p'\times p'$ identity matrix,
and $c$ is a scalar controlling the prior variance of $\bd{\phi}_h$.
The parameter 
$\alpha_{hi}$ links the survival process for cognitive outcomes with the longitudinal biomarker model.
The residual survival time after onset,
$T_i - \alpha_i$, follows a Weibull distribution with scale $\lambda_{hi}$. In practice, we often do not observe time $T_i$ due to right censoring, rather we observe the last visit for a subject which is recorded as a censoring time $C_i$. To accommodate this, we adopt a data augmentation approach that considers the unobserved event time $T_i$ as missing and the value is imputed during posterior sampling with a lower bound defined by $C_i$ as described in \citet{nimble-manual2026}.

\subsection{Posterior inference through Markov Chain Monte Carlo}\label{sec3_2_mcmc}

\subsubsection{Sampler implementation}\label{sec3_2_1_imp}

Our model is implemented in the \texttt{NIMBLE} library \citep{nimble-software2026} in \texttt{R 4.5.3}, which allows flexible specification of the hierarchical structure and customized Markov chain Monte Carlo (MCMC) sampling strategy. 
We define a three-parameter Weibull distribution as a user-defined distribution within the \texttt{NIMBLE} framework.
We construct the model using \texttt{nimbleModel} and compile with \texttt{compileNimble} prior to MCMC execution. We manually modify the default sampling scheme for several parameters: 
we implement six distinct \texttt{AF\_slice} samplers,
corresponding to 
covariate effects $\bd\gamma_h$, 
smoothing variance parameters $\sigma^2_{\theta_h}$,
spline coefficient vectors for the first biomarker $\bd\theta_1$, 
and the second biomarker $\bd\theta_2$, 
the Weibull regression coefficients $\bd\phi_h$, 
and the time shift parameters $\alpha_{hi}$. This customized sampling design is used to improve mixing and stability for correlated parameters involved in nonlinear model components and constrained likelihood structures. To restrict $d_{hij}$ to satisfy the constraint
$\mathbb{I}(L_h < d_{hij} < U_h)$ introduced in Section~\ref{sub2sec-biom}, 
we impose the following bounds on the sampling distribution of $\alpha_{hi}$:
$\alpha_{hi} \in ( \max_j t_{hij} - U_h, \, \min_j t_{hij} - L_h ),$
which guarantees that 
$d_{hij} = t_{hij} - \alpha_{hi} \in (L_h, U_h) \quad \forall j = 1, \dots, J_{hi}.$
This bounded sampling strategy is implemented using
\texttt{dconstraint} within \texttt{NIMBLE} which effectively rejects proposals for $\alpha_{hi}$ that violate the linear constraint above. Finally, chains are initialized to accommodate the above constraints, with further details provided in Sections \ref{sub2sec-imp} and \ref{subsec-data-implementation} for the simulation and real data analysis, respectively. 

\subsubsection{Posterior inference}
Posterior inference is performed using MCMC as implemented in the \texttt{NIMBLE} \citep{nimble-software2026} library in \texttt{R}. 
Trace plots and autocorrelation diagnostics are also visually inspected for all parameters. Posterior summaries are obtained using posterior means and 95\% credible intervals. 
The exception is for disease progression trajectories $f_h(\cdot)$, where coverage is evaluated using point-wise
intervals of the form
$\widehat f_h(d_h) \pm 1.96\,\widehat{\operatorname{sd}}\{f_h(d_h)\}$
rather than the original posterior quantile-based 95\% credible intervals where $\widehat f_h(d_h)$ is the posterior mean and $\widehat{\operatorname{sd}}\{f_h(d_h)\}$ is the posterior standard deviation.
This approximation is required for the disease progression trajectory credible intervals as posterior quantiles led to under coverage in simulations. 
One likely explanation for this underestimation of disease progression trajectory posterior variation is due to constraints on the disease progression trajectory, and this heuristic adjustment produces satisfactory performance as demonstrated in Section \ref{sec-sim}. 

\subsection{Posterior predictive distribution}
\label{subsec-3 ppd}

When biomarker data is observed for a new subject given by $\mathcal{D}^{*}$, it is often of interest to predict the vector representing their age of positivity $\bd{\alpha}^* \in \mathbb{R}^H$ given prior training data $\mathcal{D}_{\text{train}}$. Grouping all other model parameters into the vector $\bd{\varphi}$, the posterior predictive distribution is given by
\begin{gather}
p(\bd{\alpha}^* \mid \mathcal{D}^*, \mathcal{D}_{\text{train}}) = \int p(\bd{\alpha}^* \mid \mathcal{D}^*, \bd{\varphi}) \, p(\bd{\varphi} \mid \mathcal{D}_{\text{train}}) \, d\bd{\varphi},
\end{gather}
which is not available in closed form but can be sampled via MCMC. To do so, we save the MCMC chains generated during sampling from the posterior distribution $p(\bd{\varphi} \mid \mathcal{D}_{\text{train}})$, fix them as constants, and feed them into a new \texttt{NIMBLE} sampler which accepts the newly observed data $\mathcal{D}^*$ and treats $\bd{\alpha}^*$ as missing data to be imputed. The imputed $\bd{\alpha}^*$ is constrained to lie within the prespecified grid as described in Section \ref{sec3_2_1_imp}. Thus, posterior samples from (8) can be obtained with ease by leveraging the saved chains from the training data.

\section{Simulation Studies}\label{sec-sim}

We utilize an extensive set of simulations to assess the operational characteristics of our proposed B-TAJ DPM under settings similar to our motivating ADNI data. 

\subsection{Data generation}\label{sub2sec-gen}
In our simulations, we set up $H=2$ pathological biomarkers for $n = |\mathcal{I}_1 \cup \mathcal{I}_2|$ subjects with a nested design $\mathcal{I}_2 \subseteq \mathcal{I}_1$, where $|\mathcal{I}_2| = |\mathcal{I}_1|/2 = 500$.
We set $p=3$, corresponding to two demographic covariates and an intercept term. The covariates 
$\bd x_1$ and $\bd x_2$ are simulated independently from a Bernoulli$(0.5)$ distribution, and the intercept is fixed at 1.
For each subject $i$ in biomarker $h$, the number of longitudinal observations $J_{hi}$ is generated from a discrete distribution on $\{2,3,4,5,6\}$ with corresponding probabilities $\{0.44, 0.27, 0.13, 0.10, 0.06\}$ as observed in our real data.
We define a disease-time grid $\{d_\ell\}_{\ell=1}^{n_d}$, with $n_d = 250$ and spacing $\Delta d = 0.25$,
and biomarker positivity thresholds $\tau_h = 1$.
Then we construct a common cubic B-spline basis with intercept for both pathological markers 
with the defined disease-time grid, 
yielding a design matrix $\mb B \in \mathbb{R}^{n_d \times K}$. We place $K = 10$ equally spaced knot locations 
$\kappa_1 < \kappa_2 < \cdots < \kappa_{10}$ over $[L_h,U_h]$, where 
$\kappa_1 = L_h$ and $\kappa_{10} = U_h$ are treated as boundary knots. 
The spline coefficients $\bd\theta_h \in \mathbb{R}^{K}$ 
are generated to induce a disease progression trajectory with an increasing rate of change followed by a relatively flat segment.
For the $i$-th subject in the $h$-th marker, 
we generated $(t_{hij}, y_{hij})$ for $h \in \{1, 2\}$ as: 
\begin{gather}
    y_{hij} = \int_{L_h}^{U_h} \exp\{\mb B(d_h)' \bd\theta_h \}\,dd_h + \epsilon_{hij}, \;
     \epsilon_{hij} \sim  \mathcal{N}(0, \sigma^2_{\epsilon_h}) \\
    t_{hij} = d_{hij} + \alpha_{hi}, \
    \bd\alpha_i \sim \mathcal{BVN}(\bd\mu_i, \mb\Sigma_\alpha ), \\
  \mu_{hi} = \bd{x}_i'\bd{\gamma}_h,\ \bd\gamma_{h} \sim \mathcal{N}\{
    (\mu_{h\gamma},\bd{0}_{p-1})', \bd b\mb{I}_p\},  \label{eq:mvt2}
\end{gather}
where $\mathcal{BVN}(\cdot)$ is a bivariate normal distribution with mean $\bd\mu_i = (\mu_{i1}, \mu_{i2})'$, 
$\mb\Sigma_\alpha  = 
\begin{bmatrix}
\sigma_{\alpha_1}^2 \quad \rho\sigma_{\alpha_1}\sigma_{\alpha_2}\\
\rho\sigma_{\alpha_1}\sigma_{\alpha_2} \quad \sigma_{\alpha_2}^2
\end{bmatrix}
$ is the $2 \times 2$ covariance matrix, where $\rho$ is the correlation coefficient and $\sigma_{\alpha_1} = \sigma_{\alpha_2}$, 
and $\bd x_i \in \mathbb{R}^p$.
We generate the intercept coefficients with 
$\mu_{1\gamma} = 50$, $\mu_{2\gamma} = 55$, 
and $\bd b = (1,10,10)$.
We vary $\sigma_{\epsilon_h}$ and $\sigma_{\alpha_h}$ to assess the robustness of the simulated data. We conducted simulations under combinations of 
two observation noises $\sigma_{\epsilon_h}$  = 0.05 and 0.10, 
and two time shift noises $\sigma_{\alpha_h}$  = 4 and 8,
yielding a total of four settings. 

For the survival component, 
we first identified, 
among subjects with the second biomarker, 
the earliest event time associated with an observation exceeding the prespecified threshold value of 1. 
For every eligible subject, 
we retained the minimum simulated event time among observations satisfying $y_{2ij} \geq 1$, 
thereby defining the subject-specific event time for the survival analysis.
For subjects with the second biomarker, 
we generated a time-to-event outcome from a three-parameter Weibull distribution:
\begin{gather}
    T_i \sim \text{Weibull}_3(\nu, \lambda_{i}, \alpha_{i2}), \\
    \lambda_i = \exp(\phi_1 + x_{i2}' \phi_2 + \alpha_{i2} \phi_3),
\end{gather}
where $\nu = 1.84, \phi_1 = 4.765, \phi_2 = 0.062, \phi_3 = -0.039$ are fixed parameters
in order to simulate time distribution observed in ADNI dataset, 
$\lambda_{i}$ is the scale parameter, 
$\alpha_{i2}$ is the time shift for the second biomarker,
and $x_{i2}$ is the second covariate.
Independent right censoring is imposed by generating a censoring indicator from Bernoulli$(0.3)$. 
Subjects with censoring indicator equal to 1 were treated as right-censored, 
whereas the remaining subjects were treated as having an observed event. 
Let $T_i$ denote the true event time. 
For censored subjects, the event time was recorded as missing and the corresponding censoring time was set to $C_i = T_i$. 
For uncensored subjects, the observed event time was retained and the censoring time was set to infinity. 
Thus, the observed data follow the conventional representation $(\tilde{T}_i, C_i)$, where
\[
\tilde{T}_i =
\begin{cases}
T_i, & \text{if the event is observed},\\
\text{NA}, & \text{if right-censored},
\end{cases}
\quad
C_i =
\begin{cases}
+\infty, & \text{if the event is observed},\\
T_i, & \text{if right-censored}.
\end{cases}
\]
This construction yields approximately 30\% right censoring in the simulated survival outcomes.

\subsection{Evaluation metrics}\label{sub2sec-eval}

Performance of the proposed estimation procedure is assessed using 
normalized mean squared error (NMSE) and relative squared error (RSE), defined in terms of the norms of the deviations between the estimates and the corresponding target quantities. 
Specifically, we utilize NMSE, 
NMSE $=\|\hat{\theta}-\theta\|_2^2/\|\theta\|_2^2$, 
on time shift ($\bd\alpha_h$), 
covariate coefficients ($\bd\gamma_h$),
time shift noise ($\sigma_h$),
observation noise ($\sigma_{\epsilon_h}$), covariate coefficients ($\bd\phi$) for three-parameter Weibull distribution.
To evaluate the performance of the proposed estimation for the progression curves ($f_h(\cdot)$),
we utilize RSE, $\text{RSE}=\|\hat{f}(s) - f(s)\|^2/\|f(s)\|^2$, for one-dimensional functional components,
where $\|f(s)\|^2 = \int f^2(s)ds$.
We also estimate coverage probability of the 95\% credible intervals by 
computing the proportion of times that the nominal 95\% interval contained the corresponding true value.
For scalars, coverage is calculated as the average indicator that the true parameter value falls within its estimated 95\% posterior credible interval. 
For the progression trajectories $f_h(\cdot)$, coverage is evaluated point-wise on the fixed disease-time grid.
The coverage probabilities of the $95\%$ credible intervals are evaluated for the time-shift parameters ($\sigma_h$), the progression curves ($f_h(\cdot)$), and the regression coefficients in both the time-shift ($\bd\gamma_h$) and Weibull models ($\bd\phi$). 

\subsection{Implementation details}\label{sub2sec-imp}

We obtain posterior inference as described in Section \ref{sec3_2_mcmc}. In simulation studies, we use $5,000$ total iterations with $3,000$ iterations discarded as burn-in. After a burn-in period, posterior samples were collected. 
For regression coefficients, we assume $\mu_{1\gamma} = \mu_{2\gamma} = 50$ and $a = b = 10$, and $c = 100$. For the age of positivity variance parameters, we assume $\zeta_h = 5$ and $\eta_h =2$ for $h =1, 2$.
Several parameters are initialized to satisfy the support constraints of the hierarchical model. 
The spline coefficient vectors for amyloid and tau are initialized using an evenly spaced sequence from $-9$ to $-3$.
This initialization generates a small but increasing initial rate function and provides a stable initial trajectory for both biomarkers. The subject-specific positivity ages $\alpha_{hi}$ are initialized from the observed distributions. For each subject, the initial values of $\alpha_{hi}^{(0)}$ are then set using the subject-specific mean calendar age ${\bar t_{hi}} = \sum_{j=1}^{J_{hi}} t_{hij}/{J_{hi}}$. 
This initialization places the latent disease time
$d_{hij} = t_{hij} - \alpha_{hi}$ within the bounds defined in Section \ref{sub2sec-dpm}.
We also adjust the initial biomarker positivity ages for subjects in the survival subset. 
Specifically, the initial value of $\alpha_{2i}$ is further truncated so that it is at least one year before the initialized event or censoring time:
$
\min(
\alpha_{i2}^{(0)},
T_i^{(0)} - 1),
$
which ensures that the initial values are compatible with the support of the three-parameter Weibull distribution and avoid invalid starting states for the MCMC sampler. For the survival component, event times are initialized according to censoring status. The Weibull shape parameter is initialized at 1, and the Weibull regression coefficients are initialized using the log median observed event time for the intercept and 0 for the remaining coefficients.


\subsection{Results}\label{sub2sec-res}
\begin{table}[htbp]
\caption{Performance of our model in simulation studies. Rows 1-9 report NMSE values and Rows 10-11 report the RSE values with 95\% credible intervals in parentheses.}
\label{tab:simures}
\centering
\begin{tabular}{lcccc}
\toprule
 & \multicolumn{2}{c}{$\sigma_{\epsilon_h} = 0.05$} & \multicolumn{2}{c}{$\sigma_{\epsilon_h} = 0.10$} \\
\cmidrule(lr){2-3} \cmidrule(lr){4-5}
 & $\sigma_{\alpha_h}  = 4$ & $\sigma_{\alpha_h}  = 8$ & $\sigma_{\alpha_h}  = 4$ & $\sigma_{\alpha_h}  = 8$ \\
\midrule\noalign{}
$\bd\alpha_1$    & $0.013\ (94.27)$ & $0.015\ (93.73)$ & $0.023\ (93.00)$ & $0.021\ (94.48)$ \\[-10pt]
$\bd\alpha_2$    & $0.005\ (94.27)$ & $0.005\ (94.16)$ & $0.007\ (93.22)$ & $0.009\ (94.60)$ \\[-10pt]
$\bd\gamma_1(\times10^{-4})$    & $0.269\ (95.67)$ & $0.862\ (93.67)$ & $0.516\ (95.00)$ & $1.113\ (95.00)$ \\[-10pt]
$\bd\gamma_2(\times10^{-4})$    & $0.503\ (96.00)$ & $1.107\ (96.33)$ & $0.806\ (95.00)$ & $1.504\ (94.67)$ \\[-10pt]
$\bd\phi$        & $0.024\ (88.33)$ & $0.011\ (88.67)$ & $0.029\ (91.33)$ & $0.010\ (88.33)$ \\
$\sigma_{\alpha_1}$              & 0.003 (96.00) & 0.004 (97.00) & 0.005 (96.00) & 0.006 (97.00) \\[-10pt]
$\sigma_{\alpha_2}$              & 0.006 (96.00) & 0.007 (96.00) & 0.009 (99.00) & 0.009 (93.00) \\[-10pt]

$\sigma_{1}(\times10^{-5})$      & 2.347 (80.00) & 1.669 (88.00) & 3.601 (86.00) & 4.128 (84.00) \\[-10pt]
$\sigma_{2}(\times10^{-5})$      & 2.837 (90.00) & 3.700 (86.00) & 5.508 (94.00) & 6.686 (87.00) \\

$f_1(\cdot)(\times10^{-4})$    & 0.461 (99.04) & 0.886 (97.36) & 0.827 (99.79) & 1.262 (99.37) \\[-10pt]
$f_2(\cdot)(\times10^{-4})$    & 0.587 (99.92) & 0.912 (99.24) & 1.127 (99.76) & 1.678 (99.65) \\
\bottomrule
\end{tabular}
\end{table}
Row 1-9 of Table~\ref{tab:simures} summarizes the NMSE values
across 100 simulation replications. 
Overall, the estimation accuracy decreases as either measurement noise $\sigma_{\epsilon_h}$ or the variability in the time shift $\sigma_{\alpha_h}$ increases. 
For the time-shift parameters $\bd\alpha_{h}$ the smallest NMSE values are observed under the low-noise setting $(\sigma_{\epsilon_h}=0.05,\sigma_h=4)$, with NMSEs of $0.013$ and $0.005$, respectively. 
In contrast, the largest NMSE values occur under the high-noise setting $(\sigma_{\epsilon_h}=0.10,\sigma_h=8)$, where the corresponding NMSEs increase to $0.021$ and $0.009$. 
Moreover, comparing the settings $(\sigma_{\epsilon_h}=0.10,\sigma_h=4)$ and $(\sigma_{\epsilon_h}=0.05,\sigma_h=8)$ shows that the NMSEs for $\alpha_1$ and $\alpha_2$ are larger in the former case, suggesting that measurement noise has a stronger impact on the estimation of time-shift parameters than does increased variability in the time shift.
A similar pattern is observed for the regression coefficients, $\bd\gamma_1$ and $\bd\gamma_2$. 
Their NMSE values are smallest under $(\sigma_{\epsilon_h}=0.05,\sigma_h=4)$, with values of $0.269 \times 10^{-4}$ and $0.503 \times 10^{-4}$, respectively, and largest under $(\sigma_{\epsilon_h}=0.10,\sigma_h=8)$, with values of $1.113 \times 10^{-4}$ and $1.504 \times 10^{-4}$. 
The regression coefficient estimates appear to be  sensitive to both increases in $\sigma_{\alpha_h}$ and $\sigma_{\epsilon_h}$, 
indicating that greater variability in both the time shift and measurement noise variance has an effect on the estimation accuracy of $\bd\gamma_1$ and $\bd\gamma_2$.
For the Weibull distribution parameters $\bd\phi$, the NMSE values remain relatively small across all simulation settings. 
Specifically, the NMSEs for $\bd\phi$ range from $0.080$ to $0.229$. These results suggest that the estimation of the Weibull parameters is robust to changes in both the measurement error of the biomarker, $\sigma_{\epsilon_h}$, and the variability in the time shift, $\sigma_h$.
The noise parameters $\sigma_{\alpha_h}$ and $\sigma_h$ are estimated accurately across all settings. 
For $\sigma_{\alpha_1}$ and $\sigma_{\alpha_2}$, the NMSEs increase moderately as $\sigma_{\epsilon_h}$ or $\sigma_h$ increases, suggesting some sensitivity to higher SNR. 
The NMSEs for $\sigma_1$ and $\sigma_2$ remain on the order of $10^{-6}$, indicating accurate estimation.

Row 10-11 of Table~\ref{tab:simures}
displays the RSEs for the progression trajectories, $f_1(\cdot)$ and $f_2(\cdot)$, on the order of $10^{-4}$.
The errors are comparatively small across all simulation settings, indicating that the proposed model provides accurate estimation of the underlying progression trajectories. 
For both $f_1(\cdot)$ and $f_2(\cdot)$, the RSE values increase as either $\sigma_{\epsilon_h}$ or $\sigma_h$ increases, with the smallest errors observed under $(\sigma_{\epsilon_h}=0.05,\sigma_h=4)$ and the largest errors under $(\sigma_{\epsilon_h}=0.10,\sigma_h=8)$. We also evaluate the coverage of $95\%$ credible intervals across the simulation settings in 
Table~\ref{tab:simures}. 
The results show that the coverage for the time-shift parameters, $\alpha_1$ and $\alpha_2$, and the regression coefficients, $\bd{\gamma}_1$ and $\bd{\gamma}_2$, are close to the nominal level of $95\%$, indicating reliable uncertainty quantification for these parameters. 
The coverage for the progression trajectories, $f_1(\cdot)$ and $f_2(\cdot)$, remains stable across all settings.
For the Weibull parameter $\bd{\phi}$, the coverage probabilities range from $88.33\%$ to $91.33\%$, suggesting mildly conservative or slightly lower-than-nominal interval performance in some cases.

\section{ADNI Data Analysis}

\subsection{Data Processing and Structure}
We apply B-TAJ DPM and SILA \citep{Betthauser2022} to the ADNI data. 
Amyloid observations greater than 168.5 CL and tau observations greater than SUVR greater than 3 were excluded due to data sparsity. A binary cognitive impairment indicator was defined as 1 when CDR $\ge0.5$ and 0 otherwise. APOE-$\varepsilon$4 carrier status was coded as 1 if the APOE genotype contained allele 4 and 0 otherwise. Education was dichotomized as 1 for college education or higher versus 0 for less than college. Biological sex was represented as 1 for female and 0 for male. The final covariate matrix included an intercept, centered age, APOE-$\varepsilon$4 carrier status, college education, female sex, and an APOE-$\varepsilon$4-by-female interaction term. We define a disease-time grid $\{d_\ell\}_{\ell=1}^{n_d}$, with $n_d = 201$ and spacing $\Delta d = 0.25$,
and biomarker positivity thresholds A+ = 24 CL and T+ = 1.41 SUVR. Then we construct a common cubic B-spline basis with intercept for both pathological markers with the defined disease-time grid, 
yielding a design matrix $\mb B \in \mathbb{R}^{n_d \times K}$. We place $K = 10$ equally spaced knot locations 
$\kappa_1 < \kappa_2 < \cdots < \kappa_{10}$ over $[-20\text{ years}, 30\text{ years}]$, where 
$\kappa_1 = -20$ and $\kappa_{10} = 30$ are treated as boundary knots. This positivity-age range reflects a plausible range over which amyloid and tau pathologies are expected to unfold in AD.  

\subsection{Implementation Details}\label{subsec-data-implementation}

We obtain posterior inference as described in Section \ref{sec3_2_mcmc} based on 15,000 iterations with the first 10,000 discarded as burn-in. We adopt the following prior specifications. For regression coefficients, we assume $\mu_{1\gamma} = 70$ and $\mu_{2\gamma} = 75$ and $a = 100$, $b = 10$, and $c = 100$. For the age of positivity variance parameters, we assume $\zeta_h = 5$ and $\eta_h =2$ for $h =1, 2$. We initialize certain parameters to maintain model constraints. Specifically, the spline coefficients $\bd\theta_h \in \mathbb{R}^{K}$ 
are initialized following an arithmetic sequence ranging from $-9$ to $-3$, representing a monotonic increase on the general scale of the simulated data. Initialization of the remaining parameters including the age of positivity parameters and Weibull survival parameters mimic the simulation implementation described in Section \ref{sub2sec-imp}. The real data analysis took 9.76 minutes on an Apple M3 Pro laptop.

To compare the predictive performance for age of A+ and T+ between B-TAJ DPM and SILA, we utilize five fold cross validations for amyloid and tau separately. For each biomarker in the held out fold, we restrict the analysis to individuals who have at least one observation below and one observation above the corresponding positivity threshold, thereby ensuring that a threshold-crossing event was observed. 
We further require the event to correspond to an upward crossing, from below to above the threshold, and perform linear interpolation to estimate the age of positivity. This process yields $n=116$ eligible samples for amyloid and $n=49$ eligible samples for tau.
We then perform both forward and backward prediction
for the restricted subsets: forward prediction uses only
below threshold observations, whereas backward prediction uses only above threshold
observations. To avoid using information from the other biomarker that would not be available at the prediction time, we appropriately restrict the longitudinal data used for prediction. 
Specifically, for forward prediction, we include only observations occurring no later than the maximum available pre-threshold time and for backward prediction, we include only observations occurring no earlier than the minimum available post-threshold time. Posterior predictions for B-TAJ DPM are performed as described in Section \ref{subsec-3 ppd} with the median of the posterior predictive distribution taken as an estimate for $\bd{\alpha}^*$. Predictions for SILA are performed for each biomarker using the $\mathtt{R}$ implementation of the SILA algorithm (\url{https://github.com/Betthauser-Neuro-Lab/silaR}) with a grid of 0.25 and maximum integral length of 200. Lastly, training data for SILA is restricted to subjects with at least two observations as required by their methodology. 

\subsection{Data Analysis Results}

We present the results from application of our proposed B-TAJ DPM model to the ADNI data. Figure \ref{fig:study_dta} displays subject-level trajectories for amyloid (C) and tau (D) aligned by the posterior median positivity age (years). The procedure successfully aligns subjects to a common latent positivity age timeline for each biomarker, modeling phase variation in calendar age and producing a clearer description of biomarker accumulation. The overlaid solid black lines are posterior means of the progression trajectories $f_h(d)$ for each biomarker accompanied by gray bands representing approximate 95$\%$ credible intervals. For amyloid, deposition begins to accumulate approximately 5-10 years prior to crossing the 24 CL positivity threshold, and continues at a constant rate until a slight reduction in deposition rate at 15 years post-positivity. Tau displays a longer but slower rate of deposition prior to positivity, with tau beginning to accumulate approximately 15 years prior to the 1.41 SUVR positivity threshold and accelerating to a constant rate by 10 years post positivity. Our estimated progression trajectories for amyloid and tau display similar patterns to those depicted in \citet{Heston2025}, who applied SILA to the ADNI cohort, corroborating the basic profile of amyloid and tau deposition with the added benefit of principled uncertainty quantification.

\begin{figure}[t]
\centering\includegraphics{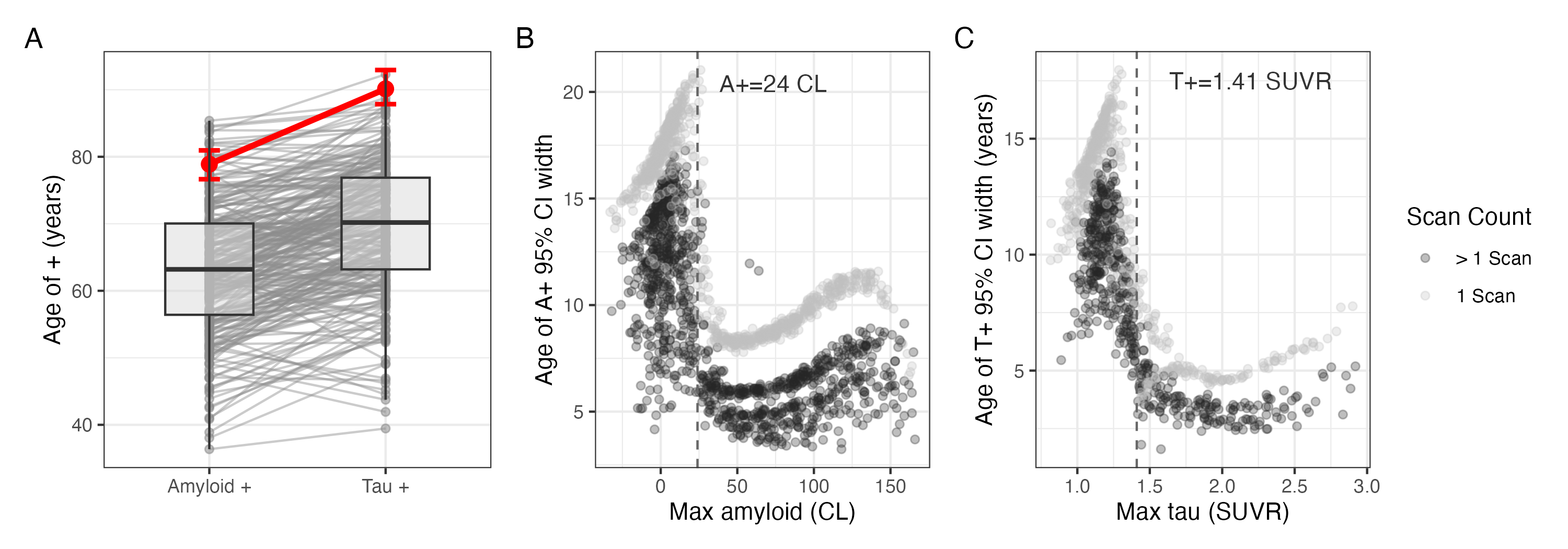}
    \caption{(A) Posterior median age of A+ and T+ presented as grey dots connected by solid lines, with a box plot superimposed for each biomarker. A single subject is highlighted in red with accompanying 95$\%$ credible intervals. Boxplots are overlaid. Note, this data represent a subset of subjects who passed both amyloid and tau positivity thresholds ($n=263$). (B, C) Subject-level 95$\%$ credible interval width for posterior age of biomarker positivity versus (B) maximum amyloid (CL) and (C) maximum tau (SUVR). Light dots correspond to subjects with one scan, whereas dark dots correspond to subjects with two or more scans. Posterior variance in age of positivity shrinks as subjects cross the positivity threshold and more scans are collected.}
\label{fig:age_uncert}
\end{figure}

Figure \ref{fig:age_uncert} (A) shows the posterior median age of A+ and T+ (grey dots connected by solid lines) among the subset of subjects who had at least one observation above the positivity threshold for both biomarkers ($n=263)$. As will be discussed below, this restriction is driven by the observation that backward prediction of age of positivity is much more accurate than forward prediction. A single subject is highlighted in red with accompanying 95$\%$ credible intervals to emphasize that B-TAJ DPM estimates the posterior distribution of age of positivity, enabling precision quantification and facilitating downstream inference. Overlaid box plots of the posterior median age of positivity show that amyloid positivity generally precedes tau positivity. However, inspection of the subject-level data shows this ordering is not always observed for individuals. By inspecting the posterior distribution for age of A+ and T+, we can enumerate the exact proportion of times the 95$\%$ CIs for age of A+ and T+ provide a definitive ordering with $n = 149$ ($54\%$) subjects experiencing A+ before T+, $n = 12$ ($5\%$) subjects experiencing T+ before A+, and  $n = 102$ ($39\%$) subjects lacking a clear ordering in age of positivity due to overlap in age of A+ and T+ 95$\%$ CIs. Continuing our discussion of uncertainty quantification for age of positivity, Figures \ref{fig:age_uncert} (B, C) display the subject-level 95$\%$ credible interval width for posterior age of biomarker positivity versus each subject's (B) maximum observed amyloid and (C) maximum observed tau measurement. Light dots correspond to subjects with one scan, whereas dark dots correspond to subjects with two or more scans. Posterior variance in age of positivity shrinks as subjects cross the positivity threshold and more scans are collected. As has been observed in prior work \citep{Betthauser2022}, it is challenging to predict age of positivity for subjects below the positivity threshold, as mirrored by the high level of model uncertainty with 95$\%$ credible interval widths ranging from 5 to 20 years. However, once an individual's maximum biomarker measurement passes this threshold, precision increases substantially. Further, the number of scans observed for a subject plays a large role in model uncertainty, with subjects who had two or more scans experiencing more precision. While this conclusion is intuitive, the observed patterns show the importance of considering uncertainty quantification when modeling subjects with heterogeneity in observed scans.

\begin{figure}[h]
\centering\includegraphics[width=0.8\textwidth]{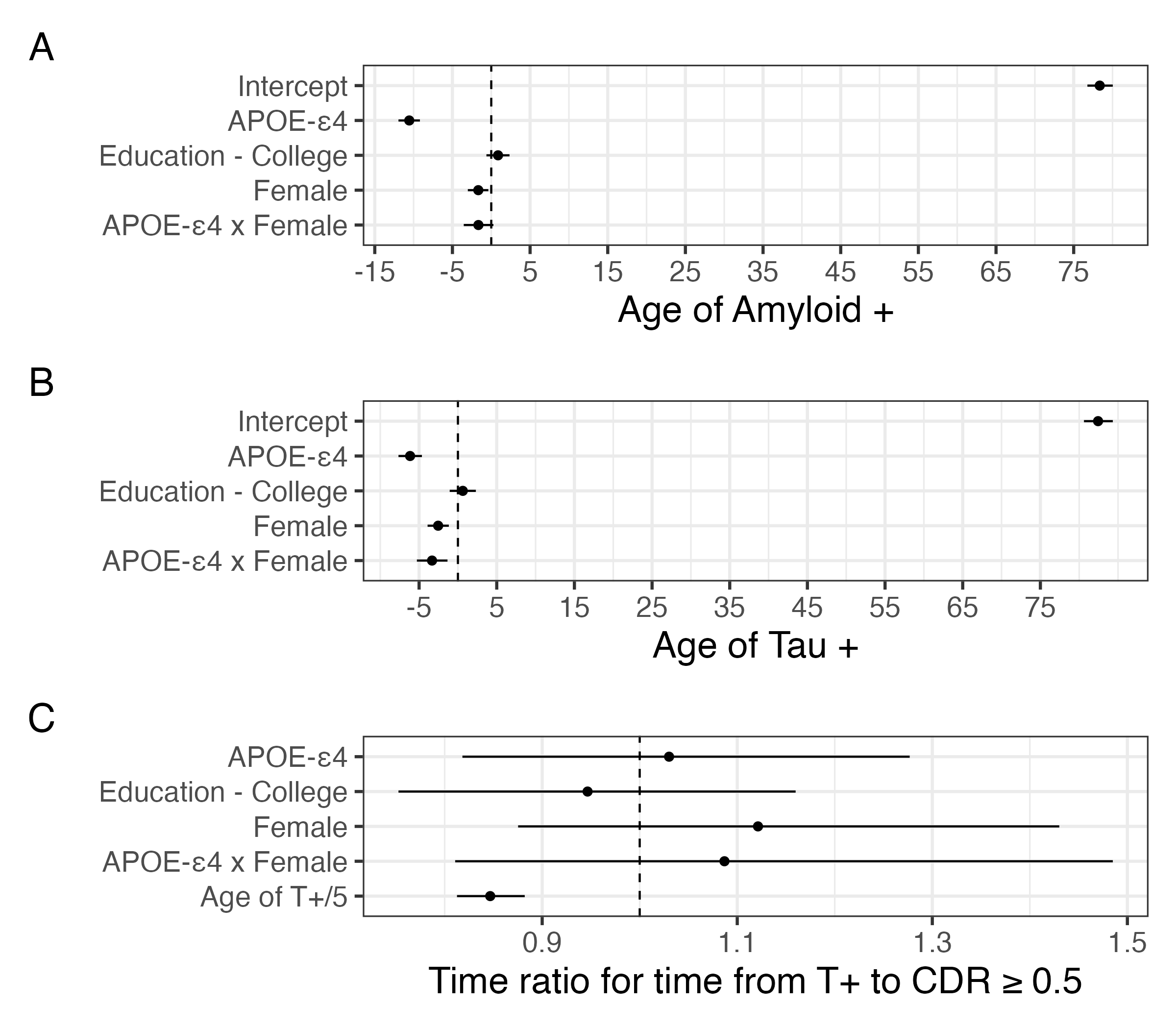}
\caption{Posterior medians and 95\% credible intervals for (A) regression coefficients $\bd{\gamma_1}$ capturing the effects of subject-level covariates on the age of A+, (B) regression coefficients $\bd{\gamma_2}$ capturing the effects of subject-level covariates on the age of T+, and (C) time ratios $\exp(\bd{\phi})$ estimating the acceleration or deceleration of time from tau positivity to CDR $\ge 0.5$.}
\label{fig:gamma}
\end{figure}

Figure \ref{fig:gamma} displays the posterior medians and 95\% credible intervals for (A) regression coefficients $\bd{\gamma_1}$ capturing the effects of subject-level covariates on the age of A+, (A) regression coefficients $\bd{\gamma_2}$ capturing the effects of subject-level covariates on the age of T+, and (C) time ratios $\exp(\bd{\phi})$ estimating the acceleration or deceleration of time from T+ to CDR $\ge0.5$. Regarding amyloid (Figure \ref{fig:gamma}(A)), in the ADNI sample a subject with biological sex male who is not college educated and not an APOE-$\varepsilon$4 carrier is expected to reach amyloid positivity at 78.8 years (95$\%$ CI: [76.8, 80.0]). Among those without APOE-$\varepsilon$4, biological sex female decreases age of A+ by 1.7 years (95$\%$ CI: [-3.0, -0.4]). For biological sex males, being an APOE-$\varepsilon$4 carrier decreases age of A+ by 10.6 years (95$\%$ CI: [-12.0, -9.4]). For biological sex females, being an APOE-$\varepsilon$4 carrier decreases age of A+ by an additional 1.6 years (95$\%$ CI: [-3.4, 0.3]). College education does not have a demonstrable effect on age of A+. With respect to tau (Figure \ref{fig:gamma}(B)), in the ADNI sample, a male participant who is not college educated and APOE-$\varepsilon$4 negative is estimated to reach tau positivity at 82.5 years (95$\%$ CI: [80.8, 84.2]). Among individuals without APOE-$\varepsilon$4, being female is associated with a reduction in age of T+ of approximately 2.6 years (95$\%$ CI: [-3.9, -1.3]). Among males, APOE-$\varepsilon$4 carriage is associated with a substantially earlier onset of tau positivity, with a decrease of approximately 6.2 years (95$\%$ CI: [-7.7, -4.7]). In females, the additional APOE-$\varepsilon$4 effect further reduces the age of T+ by about 3.3 years (95$\%$ CI: [-5.2, -1.2]). College education does not show a meaningful association with age of A+, with credible intervals spanning zero. 

We now consider what factors can predict time from T+ to symptoms of mild cognitive impairment (Figure \ref{fig:gamma}(C)), and see that for every five year increase in age of T+, the time from T+ to CDR $\ge 0.5$ decreases by a factor of 0.85 (95$\%$ CI: [0.81, 0.88]). This suggests that the progression from tau positivity to cognitive decline occurs more quickly at later stages in life. None of the other factors (biological sex, APOE-$\varepsilon$4, or college education) are associated with progression from tau positivity to cognitive impairment after the effect of age of T+ has been controlled for.

Lastly, Table \ref{tab:pred_realdata} shows forward and backward predictive accuracy for age of A+ and T+ comparing B-TAJ DPM and SILA. Results are produced via five fold cross validation with predictions assessed using mean absolute errors (MAE) and Pearson correlation coefficients. A general trend across methods is that forward prediction produces greater errors than backward prediction. With the exception of forward prediction of amyloid, our proposed B-TAJ DPM model produces lower MAE and higher correlation coefficients than SILA. The reduction in MAE in most cases is almost half a year. These results demonstrate that our B-TAJ DPM not only has higher predictive accuracy than SILA while simultaneously providing uncertainty quantification via the posterior predictive distribution sampled as described in Section \ref{subsec-3 ppd}.

\begin{table}[h]
\renewcommand{\arraystretch}{0.75}
\caption{Forward and backward predictive accuracy for age of A+ and T+ comparing B-TAJ DPM and SILA. Results are produced via five fold cross validation with predictions assessed using mean absolute errors (MAE) and Pearson correlation coefficients.}
\centering
\begin{tabular}{lcccc}
\toprule
 & \multicolumn{2}{c}{Amyloid (CL)} & \multicolumn{2}{c}{Tau (SUVR)} \\
\cmidrule(lr){2-3} \cmidrule(lr){4-5}
 & MAE$\downarrow$ & Correlation$\uparrow$ & MAE$\downarrow$ & Correlation$\uparrow$ \\
\midrule\noalign{}
Forward 
SILA & 1.744 & 0.942 & 1.770 & 0.967 \\
Forward B-TAJ DPM & 2.014 & 0.951 & 1.147 & 0.980 \\
Backward 
SILA & 1.603 & 0.956 & 1.531 & 0.966 \\
Backward B-TAJ DPM & 1.173 & 0.975 & 1.050 & 0.985 \\
\bottomrule
\end{tabular} \\
Note: $\uparrow$ means higher is better, $\downarrow$ means lower is better.
\label{tab:pred_realdata}
\end{table}

\section{Conclusion}\label{sec-conc}

In this paper, we introduced the Bayesian Threshold-Aligned Joint Disease Progression Model (B-TAJ DPM), a novel generative, semi-parametric hierarchical framework designed to model multivariate, monotonic biomarker trajectories while simultaneously mapping them to a cognitive survival endpoint. By anchoring individual latent disease timelines at outcome-specific positivity thresholds and integrating an accelerated failure time model, B-TAJ DPM effectively addresses critical limitations of traditional ND-DPMs while simultaneously providing principled uncertainty quantification. Our extensive simulation studies validated the model's operational characteristics. When applied to the ADNI data, the B-TAJ DPM provided several insights. First, subject-level variation in age of positivity was partitioned across known risk factors while simultaneously quantifying how individuals differ from these expected trends. Key indicators of uncertainty in age of positivity estimates were maximum observed biomarker levels and number of imaging scans available. Second, our framework revealed that an older age of tau positivity significantly compresses the subsequent window from tau positivity to mild cognitive impairment. Third, B-TAJ DPM consistently outperformed a popular existing approach for predicting age of positivity in held out samples. Future work will focus on expanding the B-TAJ DPM framework to accommodate subject heterogeneity in the rate of pathological progression, allowing for a deeper interrogation of variation not only in the age of positivity but also the speed at which progression occurs.

\section{Acknowledgments}\label{disclosure-statement}


Data collection and sharing for the Alzheimer's Disease Neuroimaging Initiative (ADNI) is funded by the National Institute on Aging (National Institutes of Health Grant U19AG024904). The grantee organization is the Northern California Institute for Research and Education. In the past, ADNI has also received funding from the National Institute of Biomedical Imaging and Bioengineering, the Canadian Institutes of Health Research, and private sector contributions through the Foundation for the National Institutes of Health (FNhi) including generous contributions from the following: AbbVie, Alzheimer's Association; Alzheimer's Drug Discovery Foundation; Araclon Biotech; BioClinica, Inc.; Biogen; Bristol-Myers Squibb Company; CereSpir, Inc.; Cogstate; Eisai Inc.; Elan Pharmaceuticals, Inc.; Eli Lilly and Company; EuroImmun; F. Hoffmann-La Roche Ltd and its affiliated company Genentech, Inc.; Fujirebio; GE Healthcare; IXICO Ltd.; Janssen Alzheimer Immunotherapy Research \& Development, LLC.; Johnson \& Johnson Pharmaceutical Research \& Development LLC.; Lumosity; Lundbeck; Merck \& Co., Inc.; Meso Scale Diagnostics, LLC.; NeuroRx Research; Neurotrack Technologies; Novartis Pharmaceuticals Corporation; Pfizer Inc.; Piramal Imaging; Servier; Takeda Pharmaceutical Company; and Transition Therapeutics.

\section{Data Availability Statement}\label{data-availability-statement}

The ADNI data is available upon request from \url{https://adni.loni.usc.edu/}.

\section{Generative Artificial Intelligence Acknowledgment}

During the preparation of this manuscript, the author(s) used Google Gemini (3.5 Flash) to assist with minor text adjustments. Specifically, the large language model was utilized to provide small editorial suggestions, improve conciseness, and refine phrasing for short fragments of text. The tool was used solely to enhance the readability and clarity of the author-generated text; it was not used to create new scientific ideas, interpret data, or synthesize conclusions. The author(s) reviewed, edited, and approved all final suggestions and take full responsibility for the content of the manuscript.








\bibliography{references}

\end{document}